\newcommand{\be}{\begin{equation}}
\newcommand{\ee}{\end{equation}}
\newcommand{\bea}{\begin{eqnarray}}
\newcommand{\eea}{\end{eqnarray}}
\newcommand{\bean}{\begin{eqnarray*}}

\newcommand{\eean}{\end{eqnarray*}}
\font\upright=cmu10 scaled\magstep1
\font\sans=cmss10
\newcommand{\ssf}{\sans}
\newcommand{\stroke}{\vrule height8pt width0.4pt depth-0.1pt}
\newcommand{\Z}{\hbox{\upright\rlap{\ssf Z}\kern 2.7pt {\ssf Z}}}

\newcommand{\C}{{\rlap{\rlap{C}\kern 3.8pt\stroke}\phantom{C}}}
\newcommand{\R}{\hbox{\upright\rlap{I}\kern 1.7pt R}}
\newcommand{\CP}{\C{\upright\rlap{I}\kern 1.5pt P}}
\newcommand{\PP}{\hbox{\upright\rlap{I}\kern 1.5pt P}}

\newcommand{\identity}{{\upright\rlap{1}\kern 2.0pt 1}}

\newcommand{\pp}{\Delta}
\newcommand{\HH}{\mbox{\hbox{\upright\rlap{I}\kern 1.7pt H}}}

\newcommand{\fr}{\frac}
\newcommand{\lm}{\lambda}
\newcommand{\ra}{\rightarrow}

\newcommand{\pr}{\partial}
\newcommand{\hs}{\hspace{5mm}}

\newcommand{\dg}{\dagger}

\newcommand{\ve}{\varepsilon}
\newcommand{\acc}{\\[3mm]}

\newcommand{\zb}{{\bar z}}
\newcommand{\ub}{{\bar u}}
\newcommand{\fac}{(1+\vert z\vert^2)^2}
\input{epsf}

\documentstyle[12pt,a4wide]{article}

\newcommand{\news}{\setcounter{equation}{0}}
\begin{document}
\title{\vskip -90pt
\begin{flushright}
\end{flushright}\vskip 30pt
{\bf \large \bf $SU(N)$ Skyrmions from Instantons\footnote{To
appear in Nonlinearity}}\\[30pt]
\author{Theodora Ioannidou\\[10pt]
\\{\normalsize  {\sl Institute of Mathematics, University of Kent at
 Canterbury,}}\\
{\normalsize {\sl Canterbury, CT2 7NF, U.K.}}\\
{\normalsize{\sl Email : T.Ioannidou@ukc.ac.uk}}\\}}
\date{}
\maketitle

\begin{abstract}
Atiyah and Manton \cite{AM} have outlined a scheme to obtain 
approximations to the $SU(2)$ skyrmions from instantons in 
 $\R^4$.
In this paper we apply this  scheme  to
construct, in an explicit form, approximations to static spherically 
symmetric $SU(N)$ skyrmions  with various baryon numbers.
In particular we show how to obtain the skyrmions from instantons using
harmonic maps into complex projective spaces.
\end{abstract}

\section{Introduction}
\news\ \ \ \ \ \
In this paper we construct a class of cylindrically symmetric $SU(N)$
instantons and calculate some of their properties, like the topological charges.
This construction involves harmonic maps of the plane into \CP$^{N-1}$.
The symmetry of the solutions determines 
the dependence of the fields on the three dimensional polar angles and leaves
unknown only the dependence on the three dimensional radius ($r$) and the 
Euclidean time $(\tau)$.
These solutions describe instantons with the same spatial location 
 but centered at different times and with different scales.

Atiyah and Manton \cite{AM} have observed that computing the holonomy of
$SU(2)$ instantons in $\R^4$ generates configurations in $\R^3$ which are
good approximations to solutions of the Skyrme model.
Here, we extend their construction to $SU(N)$ and
derive explicitly some $SU(3)$ spherically symmetric skyrmions from the
$SU(3)$ cylindrically symmetric instantons.
In addition, we connect these skyrmion solutions with the ones  
obtained using the harmonic maps of $S^2$ to \CP$^{N-1}$ \cite{IPZ}.

Perhaps we should make it clear that there are several approaches  to
studying cylindrically symmetric instantons and spherically symmetric
skyrmions. 
The main aim of this paper is not the construction of new instanton or
skyrmion solutions, but rather to gain a better understanding of the
correspondence between them and harmonic maps.
In particular, using the Atiyah-Manton procedure \cite{AM}, 
we obtain explicit closed form approximations for the profile functions 
of the $SU(N)$ Skyrme fields, which until now could only be determined 
numerically.

\section{$SU(N)$ Instantons}
\news\ \ \ \ \ \
Instantons are solutions of the $SU(N)$ Yang-Mills equations 
in $\R^4$ which are derived from the action functional
\be
S=-\frac{1}{16\pi^2}\int \mbox{tr}(F_{ij}^2)
\ d^4x
\label{ac}
\ee
which is expressed in terms of topological charge units, ie $S\geq |k|$.
Here $k$ is the topological charge 
(or Pontryagin index) which counts the number of instantons 
of the configuration.
[Since we are in the self-dual sector, ie $F=*F$, the topological charge 
$k$ is given by (\ref{ac})].
$A_i$, for $i=0,1,2,3$, is the $su(N)$-valued gauge potential, with
field strength $F_{ij}=\partial_iA_j-\partial_jA_i+[A_i,A_j]$ and
covariant derivative $D_i=\pr_i+[A_i,]$.
The  finiteness of the action implies that the field strength must go to zero
at  spatial infinity, which means that the gauge field $A_i$ 
must be a pure gauge at spatial infinity.

Variation of the action (\ref{ac}) gives the second order 
Yang-Mills equations
\be
D_iF_{ij}=0.
\label{ymh}
\ee

The starting point for our investigation is the introduction of the 
coordinates $u,\ub,z,\zb$ on $\R^4.$
In terms of the usual spherical coordinates $r,\theta,\varphi$  the
Riemann sphere variable is $z=e^{i\varphi}\tan(\theta/2)$, while $u=r+i\tau$.
Using these coordinates the Yang-Mills equations (\ref{ymh}) 
take the form
\bea
D_u((u+\ub)^2 F_{u\ub})+\fac(D_z F_{u{\zb}}+D_{\zb} F_{uz})&=&0\label{ymh1}\\
D_\ub((u+\ub)^2 F_{u\ub})-\fac(D_z F_{\ub {\zb}}+D_{\zb}
F_{\ub z})&=&0\label{ymh2}\\
D_u F_{\ub z}+D_\ub F_{uz}-\fr{1}{(u+\ub)^2}
D_z({\fac} F_{z{\zb}})&=&0\label{ymh3}\\
D_u F_{\ub \zb}+D_\ub F_{u \zb}+\fr{1}{(u+\ub)^2}D_\zb({\fac}F_{z{\zb}})
&=&0
\label{ymh4}
\eea
while the corresponding action (\ref{ac}) becomes
\be
 S\!=\!-\fr{1}{8\pi^2}\int\!\mbox{tr}\left(-4F_{u
\ub}^2+\fr{8\fac}{(u+\ub)^2}(|F_{uz}|^2+|F_{u\zb}|^2)
-\fr{4(1+|z|^2)^4}{(u+\ub)^4}F_{z\zb}^2\right) 
r^2dr\,d\tau\,\fr{2idz d\zb}{ (1+|z|^2)^{2}}.
\ee
A complex gauge can, at least for the self-dual case, be chosen so that
\be
A_\ub=0,\hs \hs \hs A_\zb=0.
\label{Gauge}
\ee
The full equations (\ref{ymh2}) and (\ref{ymh4}) become then
the total $\pr_\ub-$  and $\pr_\zb-$  derivative, respectively, of the 
self-dual Yang-Mills equation
\be
\pr_\ub A_u+\fr{\fac}{(u+\ub)^2}\, \pr_\zb A_z=0
\label{Bog}
\ee
while the other two become simply $F_{uz}=0$, ie they define the gauge
\be
A_z=H^{-1}H_z,\ \ \  \ \ \ \ \ \ \ \ \ \ \ A_u=H^{-1}H_u
\label{gauge}
\ee
where $H\in SL(N,\C)$ is a Hermitian matrix and 
subscripts denote partial differentiation.

To proceed further we need to briefly recall some results about harmonic maps
of the two-dimensional \CP$^{N-1}$ sigma model.
See Zakrzewski \cite{Za} for a more detailed account of two-dimensional
sigma models and their solutions.

\subsection{Harmonic Maps}
\news\ \ \ \ \ \
The harmonic map (or sigma model) equations for the \CP$^{N-1}$ model 
are given by
\be
[P_{z\zb},P]=0  
\label{hmap}
\ee
where $P$ is an $N\times N$ Hermitian projector.
 
One set of solutions to these equations are the 
instantons given by
\be
P(f)=\frac{ff^\dagger}{\vert f\vert^2}
\label{ftop2}
\ee
where $f(z)$ is an $N$-component column vector which is a holomorphic  
function of $z$ and  whose degree is equal to the topological
charge of the sigma model. 
Another set of solutions are the anti-instantons,
which have the same form but this time $f$ is an anti-holomorphic function, 
and then the sigma model topological charge is minus the degree of $f.$

For $N=2$ these are all the finite action solutions, but for $N>2$ there are
other non-instanton solutions. 
These can be described by introducing the operator $\pp$ defined  by its 
action on any vector $f\in \C^N$ as
\be
\pp f=\pr_z f- \fr{f \,(f^\dg \,\pr_z f)}{|f|^2}
\ee
and then define further vectors $\pp^k f$ by induction:
$\pp^k f=\pp(\pp^{k-1} f)$.

To proceed further we note the following useful properties of
$\pp^k f$ when $f$ is holomorphic:
\begin{eqnarray}
\label{bbb}
&&(\pp^k f)^\dg \,\pp^l f=0, \hs \hs \hs k\neq l\acc
&&\pr_{\bar{z}}\left(\pp^k f\right)=-\pp^{k-1} f \fr{|\pp^k
f|^2}{|\pp^{k-1} f|^2},
\hs \hs
\pr_{z}\left(\fr{\pp^{k-1} f}{|\pp^{k-1} f|^2}\right)=\fr{\pp^k
f}{|\pp^{k-1}f|^2}.
\label{aaa}
\end{eqnarray}
These properties either follow directly from the definition of $\pp$
or are easy to prove \cite{Za}.
It is also convenient to define projectors $P_k$ corresponding to the
family of vectors $\pp^k f$  as
\be
P_k=P(\pp^k f), \ \ k=0,\dots,N-1.
\ee
Applying $\pp$ a total of $N-1$ times to a holomorphic vector gives an 
anti-holomorphic vector, so that a further application of $\pp$ gives the  
zero vector and hence no corresponding projector.

The projectors $P_k$ are solutions of the harmonic map equations (\ref{hmap})
and all solutions can be found in this way by starting with an appropriate
holomorphic vector $f$. 
In the \CP$^1$ case the operator $\pp$ converts a holomorphic vector 
to an anti-holomorphic vector, that is,  instantons to anti-instantons 
and these are all the solutions in this case.

Note that the projectors obtained from this sequence always satisfy 
the relation $\sum_{k=0}^{N-1}P_k=1$.

\subsection{Constructing the Instantons}
\news\ \ \ \ \ \
The self-dual Yang-Mills equation (\ref{Bog}) after the gauge choice (\ref{gauge})
 is equivalent to the single equation for $H$
\be
\pr_\ub (H^{-1} H_u)+\fr{\fac}{(u+\ub)^2}\, \pr_\zb (H^{-1} H_z)=0.
\label{ins}
\ee
This is similar to an equation introduced by Jarvis \cite{J} for studying monopoles,
under a dimensional reduction of the time $\tau$.
As we are going to illustrate, exact $SU(N)$ instanton solutions 
of (\ref{ins}) can be obtained using harmonic maps, ie 
by assuming that the field $H$ is of the form
\be
H=\exp\left\{g_0\left(P_0-\frac{1}{N}\right)
+g_1\left(P_1-\frac{1}{N}\right)+...
+g_{N-2}\left(P_{N-2}-\frac{1}{N}\right)\right\}
\label{projsun}
\ee
where $g_i=g_i(u,\ub)$ for $i=0,\dots,N-2,$ are arbitrary functions of $u$ and
$\ub$. 
Recall that the projector $P_{N-1}$ is a linear combination of the other 
projectors plus the identity matrix, which is why it is not included in
the above formula. 
The above ansatz is motivated by our recent study \cite{IS} of Bogomolny
monopoles and their construction in terms of harmonic maps.

In order for our ansatz (\ref{projsun}) to give solutions to (\ref{ins}),
the harmonic maps used must have spherical symmetry ---
 essentially the factors of $\fac$ which appear in (\ref{ins})
 must be cancelled.
The required harmonic maps are obtained by applying the above procedure
to the initial holomorphic vector
\be
f=(f_{N-1},\dots,f_j,\dots,f_0)^t, \ \ \mbox{where} 
\ \ f_j=z^j\sqrt{{N-1}\choose j}
\label{smap}
\ee
and ${N-1}\choose j$ denote the binomial coefficients.
For a discussion of the spherical symmetry of these maps see Ref. \cite{IS}.
Here we merely point out that it is at least plausible that the required
factors do indeed cancel since $\vert f\vert^2=(1+\vert z\vert^2)^{N-1}.$
We shall illustrate this explicitly in the following with some
examples.\\

{\it $SU(2)$ Case}\\

There are simplifying special cases for which we are able to perform the 
construction explicitly, the easiest example being the rational map 
$f=(z,1)^t$.
For $N=2$, there is only one profile function $g_0$ and our ansatz 
(\ref{projsun}) reduces the self-dual  equation (\ref{ins})  
to  the  following  differential equation 
\be
-(u+\ub)^2g_{0u\ub}+2(e^{g_0}-1)=0.
\label{g0}
\ee
Moreover, the topological charge $k$ is given by

\be
k=-\fr{1}{4\pi}\int\left(4g_{0u\ub}^2+\fr{4}{r^2}e^{g_0}|g_{0u}|^2
+\fr{1}{r^4} (e^{g_0}-1)^2\right)r^2dr\,d\tau
\label{enains}
\ee
where we have used the fact that  $i\int dz d\zb (1+|z|^2)^{-2}=2\pi$.

To actually solve (\ref{g0}), we first let $g_0=2\ln(\fr{u+\ub}{2})+2\rho_0$ 
(where $\rho_0$ is a new unknown function).
Then equation (\ref{g0}) becomes
\be
4\rho_{0u\ub}=e^{2\rho_0}
\ee
which is the so-called Liouville equation and can be solved explicitly using
conformal invariance. 
In this case, the function $g_0$ is
\be
g_0=2\ln\left(\fr{(u+\ub)|dh/du|}{1-|h|^2}\right).
\label{su2}
\ee
Here $h$ is, an analytic function of $u$, of the form (see Ref. \cite{Wit})
\be
h=\prod_{i=1}^{k+1}\fr{a_i-u}{\bar{a}_i+u}
\label{h}
\ee
and the $a_i$ are an arbitrary set of complex numbers (some of them perhaps
equal) constrained to have  Re $a_i>0$.
Then (\ref{su2}) provide the most general solution of (\ref{ins}) with
cylindrical symmetry and finite action.
For general $k+1$, the total multiplicity of the zeros of $h$ in the right
half plane is always $k$; therefore this solution describes $k$
instantons. 
The imaginary part of the zeroes of $h$  determines the
location of the instantons along the time axis, while the real part
determines the instanton scales \cite{Wit}.

For the special case where $h=(a_1-u)^2/(\bar{a}_1+u)^2$ (for
$a_1=\lambda+i\lambda$) the topological charge $k$ of our solution (\ref{su2})
is equal to one, ie  the configuration consists of one instanton solution
located at the origin with scale $\lambda$.\\

{\it $SU(3)$ Case}\\

For $N=3$ there are two profile functions $g_0$ and $g_1$ and
equation (\ref{ins}) reduces to
\bea
-(u+\ub)^2 g_{0u \ub}+2(e^{g_1}-1)+2(e^{g_0-g_1}-1)&=&0\nonumber\\
-(u+\ub)^2 g_{1u \ub}+4(e^{g_1}-1)-2(e^{g_0-g_1}-1)&=&0.
\label{g0g1}
\eea
It is immediately clear that there is a symmetry under the interchange of
indices $0 \leftrightarrow 1$, when applied simultaneously to $g_i$ for
$i=0,1$.
As we will show later, this symmetry can be used  to derive special 
instantons which involve a smaller number of profile functions 
and projectors.

In addition, the topological charge becomes
\begin{eqnarray}
k\!\!\!\!&=&\!\!\!\!-\fr{1}{3\pi}\!\int\{4\left(g_{0u\ub}^2+g_{1u\ub}^2
-g_{0u\ub}g_{1u\ub}
\right)
+\fr{6}{r^2}\,e^{g_0-g_1}\left(|g_{0u}|^2+|g_{1u}|^2
-g_{0u}g_{1\ub}-g_{0\ub}g_{1u}\right)+
\nonumber\acc
&&\hs\hs\fr{6}{r^2}e^{g_1}|g_{1u}|^2+\fr{1}{r^4}
\left(3e^{2(g_0-g_1)}-3e^{(g_0-g_1)}+3+3e^{2g_1}+3e^{g_1}-3e^{g_0}\right)\}
r^2dr\,d\tau\nonumber\acc
\label{duoins}
\end{eqnarray}

The profile functions equations that we obtain, ie (\ref{g0g1}),  
are related to those derived from the ansatz based approach of Bais 
et al \cite{BW} and the methods employed there can be adapted to 
solve for the functions explicitly.
For the case in which the functions $g_k$  are time independent, 
we have shown  that \cite{IS}, the equations (\ref{su3}) are precisely 
the equations for  spherically symmetric monopoles.

To solve  (\ref{g0g1}), we first let 
\bea
g_0=2\ln\left(\fr{(u+\ub)^2}{4}\right)+2\rho_0+2\rho_1\nonumber\\
g_1=2\ln\left(\fr{(u+\ub)}{2}\right)+4\rho_0-2\rho_1
\label{su3}
\eea
where $\rho_0$ and $\rho_1$ are arbitrary functions of $u$ and $\ub$.
Then  (\ref{g0g1}) simplifies to
\be
4\rho_{0u \ub}=e^{4\rho_0-2\rho_1}, \hs \hs  4\rho_{1u \ub}=e^{4\rho_1
-2\rho_0}
\label{ls}
\ee
which are a set of coupled Liouville equations.

We have two sets of solutions to (\ref{ls}).
The first one  is just the maximal embedding of $SU(2)$ in $SU(3)$ and
occurs when  $\rho_0=\rho_1$, that is for $g_0=2g_1$ (ie, using the symmetry).
Then the system (\ref{ls}) simplifies to the Liouville equation and the 
solution is just Witten's  solution \cite{Wit} embedded in $SU(3)$, 
ie $g_1$ is  given by (\ref{su2}).
In this case, the topological charge (\ref{duoins}) is exactly four times 
the $SU(2)$ one (given by (\ref{enains})).
Therefore,  for the special case $h=(a_1-u)^2/(\bar{a}_1+u)^2$, 
the configuration consists of four  instantons
--- placed on top of each other in space.

The second set of solutions to (\ref{su3}) describes an irreducible $SU(3)$
instanton for which
\bea
\rho_0=\fr{1}{2}\ln\left[\fr{3|dh/du|^2}{|h|^{1/3}(1-|h|)^2
(1+2|h|)}\right]\nonumber\acc
\rho_1=\fr{1}{2}\ln\left[\fr{3|dh/du|^2}{|h|^{2/3}(1-|h|)^2
(2+|h|)}\right]
\label{su33}
\eea
where $h$ is given by (\ref{h}).

For the special case where $h=(a_1-u)^2/(\bar{a}_1+u)^2$ 
(for $a_1=\lambda+i\lambda)$ the topological charge (\ref{duoins})
of our solution (\ref{su3}) is equal to  two.

\section{$SU(N)$ Skyrmions}
\news\ \ \ \ \ \
The Skyrme model is a nonlinear field theory which provides a good
description of low energy hardon physics.
To have finite-energy configurations, one must require that the field
$U(\vec{x},t)\in SU(N)$ goes to a constant matrix (say 1) at spatial  
infinity: $U\ra 1$ as $|\vec{x}|\ra \infty$.
This effectively compactifies the three-dimensional Euclidean space onto
$S^3$ and hence implies that the Skyrme field  can
be considered as a map from $S^3$ into $SU(N)$; and therefore it can be
classified by the third homotopy group $\pi_3(SU(N))= Z$ or,
equivalently, by the integer valued winding number 
\be
B=\fr{1}{24\pi^2}\int_{R^3} \ve_{ijk}\,\mbox{tr}\left(\pr_i U\,
U^{-1}\pr_j U\, U^{-1}
\pr_k U\, U^{-1}\right)d^3\vec{x}
\label{bar}
\ee
which is topological invariant.
This winding number classifies the solitonic sectors in the model, and
as Skyrme has argued \cite{Sk}, it may be identified with the baryon number
$B$ of the field configuration.

In the static limit, the energy of the Skyrme model is
\be
E={1\over 12\pi\sp2
} \int_{R^3}\left\{-\fr{1}{2}\,\mbox{tr}\left(\pr_iU\,
U^{-1}\right)^2-\fr{1}{16}\,\mbox{tr}\left[\pr_iU\,
U^{-1},\pr_j U\, U^{-1}\right]^2\right\}d^3\vec{x}
\label{gene}
\ee
which is expressed in the same units as the baryon number.
There is a lower bound on the energy of a given configuration in terms of the
baryon number, ie $E\geq |B|$.

Since the model is not integrable explicit skyrmion solutions are not known
and therefore, must be obtained by solving the equations numerically.
In what follows using the Atiyah-Manton approach and harmonic maps we construct
explicitly approximations to the $SU(N)$ skyrmions.

\subsection{Harmonic Maps}

Recently in \cite{IPZ}, $SU(N)$ spherically symmetric skyrmion fields  
have been constructed from the harmonic maps of $S^2$ to \CP$^{N-1}$ 
(which are {\it not} embeddings of the $SU(2)$ fields).
In fact, the Skyrme field involves the introduction of
$N-1$ projectors, ie
\be
U=\exp\left\{ig_{0_{Sk}}
\left(P_0-\fr{1}{N}\right)+ig_{1_{Sk}}\left(P_1-
\fr{1}{N}\right)+\dots+ig_{(N-2)_{Sk}}
\left(P_{N-2}-\fr{1}{N}\right)\right\}
\label{SUN}
\ee
where $g_{i_{Sk}}=g_{i_{Sk}}(r)$ for $i=0,\dots,N-2$ are the profile
functions.

Moreover, the energy (\ref{gene}) becomes
\begin{eqnarray}
E\!\!\!\!&=&\!\!\!\fr{1}{6\pi}\int\!\! r\sp2 dr\{-{1\over
N}\left(\sum_{i=0}\sp{N-2}
\dot g_{i_{Sk}}\right)\sp2+\sum_{i=0}\sp{N-2}\dot g_{i_{Sk}}\sp2 +{1\over
2r^2}\sum_{k=1}
\sp{N-1}\left(\dot g_{k_{Sk}}-\dot g_{(k-1)_{Sk}}\right)\sp2D_k
+{2\over r\sp2} \sum_{k=1}\sp{N-1}D_k\nonumber\\
&&\hspace{17mm}+{1\over 4r\sp4}
\left(D_1\sp2+\sum_{k=1}\sp{N-2}(D_k-D_{k+1})\sp2+D_{N-1}\sp2\right)\}
\label{ener}
\end{eqnarray}
where $D_k=k(N-k)(1-\cos(g_{k_{Sk}}-g_{(k-1)_{Sk}}))$.

In addition, the topological charge (\ref{bar}) takes the form
\be
B=
\fr{1}{2\pi}\sum_{i=0}\sp{N-2}(i+1)(N-i-1)\,\left(g_{i_{Sk}}-g_{(i+1)_{Sk}}
-\sin(g_{i_{Sk}}-g_{(i+1)_{Sk}})\right)_{r=0}\sp{r=\infty}.
\label{Bg}
\ee
As $g_{i_{Sk}}(\infty)=0$ (required for the finiteness
of the energy) the only contributions to the topological charge
comes from $g_i(0)$.

This way a family of exact spherically symmetric solutions of the $SU(N)$
Skyrme model has been obtained.
In fact for each $SU(N)$ model the Skyrme field involving $N-1$ projectors
leads to an exact solutions involving $N-1$ profile functions $g_{i_{Sk}}$.
These profile functions $g_{i_{Sk}}$, which satisfy $N-1$ coupled nonlinear 
ordinary differential equations and can be solved numerically, are
exhibited in Ref. \cite{IPZ}.
Next we will derive explicitly analytic forms for $g_{i_{Sk}}$ in (\ref{SUN}) 
which are good approximations to the ones obtained numerically in \cite{IPZ}.

\subsection{Derivation of Skyrmions}
\news\ \ \ \ \ \

It has been proposed by Atiyah and Manton \cite{AM} that a finite-dimensional manifold 
of Skyrme fields can be generated from  $SU(2)$ self-dual  Yang-Mills fields.
The main idea of the Atiyah-Manton scheme is to construct the holonomy
\be
U(\vec{x})={\cal T}\, \mbox{exp}\left(-\int_{-\infty}^\infty 
A_\tau(\vec{x}, \tau)\, d\tau\right)
\label{AM}
\ee
where ${\cal T}$ denotes time-ordering and $x=(\vec{x},\tau)$
denotes the time line through $\vec{x}$.

The fundamental topological result is that if we consider this holonomy as a
Skyrme field then it has baryon number $B=k$, where $k$ is the topological 
charge of the gauge potential $A_i$ (ie, instanton number).
The fields can be explicitly computed in some special cases, by integrating
simple expressions.

The gauge field $A_\tau$, using the harmonic map ansatz (\ref{projsun})
and due to the gauge choice (\ref{Gauge}), takes the form
\be
A_\tau\equiv i
A_u=-ig_{0u}(\fr{1}{N}-P_0)-ig_{1u}(\fr{1}{N}-P_1)-\dots-ig_{(N-2)u}
(\fr{1}{N}-P_{N-2}).
\label{At}
\ee
By substituting (\ref{At}) into the Atiyah-Manton formula (\ref{AM}) and 
comparing with (\ref{SUN}) 
we see that the profile functions of the instanton and Skyrme
fields are related.
In fact, the profile functions of the $SU(N)$ Skyrme fields  $g_{i_{Sk}}$ 
can be determined   analytically through the relation
\be
g_{i_{Sk}}=-\int_{-\infty}^{\infty} g_{iu}(r,\tau)\, d\tau
\label{prof}
\ee
for $i=0,\dots, N-2$.
[Recall that, these profile functions (\ref{prof}) are approximations 
to the ones  obtained  numerically in \cite{IPZ}].

Next, we will determine analytically the $g_{i_{Sk}}$ profile functions 
of the  Skyrme field, for the simplest cases of $SU(2)$ and $SU(3)$.
This way, we construct a class of spherically symmetric skyrmions and calculate
some of their properties, such as their energies and baryon numbers.
Recall that, their baryon number is equal to the topological charge of
the corresponding instanton configuration they are derived from.
\\

{\it $SU(2)$ Case}\\

There is only one profile function in this case, $g_{0_{Sk}}$, which can be
obtained from (\ref{prof}) for $g_0$ given by (\ref{su2}).
This gives the well-known one spherically symmetric skyrmion solution, with
\be
g_{0_{Sk}}=2\pi\left[1-(1+\lambda^2/r^2)^{-1/2}\right].
\ee
This Skyrme field has first been obtained in \cite{AM}.
It can be shown that the configuration consists of one skyrmion (due to
(\ref{Bg})) and the 
energy obtained from (\ref{ener}) has minimum at $\lambda^2=2.11$ and  is 
equal to $E=1.2432$, ie within 1\% of the numerically 
determined value 1.232.\\

{\it $SU(3)$ Case}\\

For $N=3$ there are two profile functions $g_{0_{Sk}}$ and  $g_{1_{Sk}}$
which again can be evaluated explicitly using (\ref{prof}).
[Recall that we have two set of solutions for the instanton functions].

When $g_0=2g_1$ for $g_1$ given by (\ref{su2}), 
we end up having one profile function for the Skyrme field 
which coincides with the profile function of a single  $SU(2)$ skyrmion, ie
\be
g_{1_{Sk}}=2\pi\left[1-(1+\lambda^2/r^2)^{-1/2}\right].
\ee
Here we note that as $g_{0_{Sk}}(0)=4\pi$ and $g_{1_{Sk}}(0)=2\pi$ the
topological charge (\ref{Bg}) of our solution is four (so is the instanton
number). 
The energy obtained from (\ref{ener}) of this configuration is exactly
four times the energy of the one $SU(2)$ skyrmion, ie $E=4\times 1.2432$.
This compares with the energy of the numerically solution 
$4 \times 1.232$  determined in  \cite{IPZ}.
[Again the minimum occurs at $\lambda^2=2.11$].

In the case where the two instanton profile functions 
 are given by equations (\ref{su3}) and (\ref{su33}), 
after performing the integration (\ref{prof}) we get
\begin{eqnarray}
g_{0_{Sk}}&=&-2\pi\left(1-\fr{\lm+3r}{2\sqrt{9\lm^2+6\lm
r+9r^2}}+\fr{\lm-3r}{2\sqrt{9\lm^2-6\lm r+9r^2}}\right)\nonumber\\
g_{1_{Sk}}&=&-\pi\left(1+\fr{\lm+3r}{\sqrt{9\lm^2+6\lm
r+9r^2}}+\fr{2(\lm-3r)}{\sqrt{9\lm^2-6\lm r+9r^2}}\right).
\end{eqnarray}
Since $g_{0_{Sk}}(0)=g_{1_{Sk}}(0)=-2\pi$ the baryon number (\ref{bar})
 is two (recall, $k=2$); 
the interpretation of  this solution is therefore that it contains two
skyrmions.
The approximate energy obtained from   (\ref{ener}) is then
$E=2.39815$ which is in good agreement with the true value $2.3764$ 
determined in \cite{IPZ}.
This solutions has, first, been obtained by Balachandran et al \cite{Bal}.

\section {Conclusion}
\news\ \ \ \ \ \

We have studied in some detail the construction of $SU(N)$ instanton from
harmonic maps. Explicit solutions have been obtained in the case of
cylindrical symmetry and we have shown how these solutions involve harmonic
maps of the plane into \CP$^{N-1}$.
In addition,  we have generate approximate solutions of the $SU(N)$ Skyrme 
model using the Atiyah-Manton procedure, ie by computing instanton holonomies.
The corresponding skyrmions are spherically symmetric, their energies are
in good agreement with their exact value and their baryon number is equal 
to the instanton number.

The same approach can be applied to generate other soliton 
approximations in lower dimensions (see for example, Ref. \cite{S}), 
leading to the fact that they might be many connections between solitons and
instantons in varying spacetime dimensions.

\section*{Acknowledgements}
\news\ \ \ \ \ \
Many thanks to Paul Sutcliffe for useful discussions and to the Nuffield
Foundation for a newly appointed lecturer award.
\\

\end{document}